\documentstyle[12pt,epsfig]{article}
    \def\lsim{\raise0.3ex\hbox{$<$\kern-0.75em\raise-1.1ex\hbox{$\sim$}}}
\def\gsim{\raise0.3ex\hbox{$>$\kern-0.75em\raise-1.1ex\hbox{$\sim$}}}
\def\noi{\noindent}
\def\nn{\nonumber}
\def\bea{\begin{eqnarray}}  \def\eea{\end{eqnarray}}
\def\beq{\begin{equation}}   \def\eeq{\end{equation}}

\def\beeq{\begin{eqnarray}} \def\eeeq{\end{eqnarray}}

\def\bmini{\setcounter{hran}{\value{equation}}
    \refstepcounter{hran}\setcounter{equation}{0}
    \renewcommand{\theequation}{\thehran\alph{equation}}\begin{eqnarray}}

\def\bminiG#1{\setcounter{hran}{\value{equation}}
\refstepcounter{hran}\setcounter{equation}{-1}
\renewcommand{\theequation}{\thehran\alph{equation}}
\refstepcounter{equation}\label{#1}\begin{eqnarray}}

\def\emini{\end{eqnarray}\relax\setcounter{equation}{\value{hran}}\ren
ewcommand{\theequation}{\thesection.\arabic{equation}}}

\def\ben{\begin{enumerate}}  \def\een{\end{enumerate}}
   \def\cite#1{[\ref{#1}]}
   \def\citd#1#2{[\ref{#1},\ref{#2}]}
   
   \def\citm#1#2{[\ref{#1}--\ref{#2}]}

\begin{document}
\begin{center}
{\Large \bf Why is the J/$\psi$ suppression enhanced at large
transverse energy ?} \\

\vskip 8 truemm
{\bf A. Capella, A. B. Kaidalov, D. Sousa}\\

Laboratoire de Physique Th\'eorique\footnote{Unit\'e Mixte de
Recherche UMR n$^{\circ}$ 8627 - CNRS}
\\ Universit\'e de Paris XI, B\^atiment 210,
F-91405 Orsay Cedex, France \\
\end{center}

\begin{abstract}
We study the ratio of $J/\psi$ over minimum bias in $Pb$ $Pb$
collisions at SPS energy. The NA50
data exhibit a sharp turn-over at $E_T \sim 100$~GeV (close to the
knee of the $E_T$
distribution) followed by a steady, steep decrease at larger
$E_T$. We show that this behaviour
can be explained by the combined effects of a small decrease of the
hadronic $E_T$ in the $J/\psi$
event sample (due to the $E_T$ taken by the $J/\psi$
trigger), together with the sharp
decrease of the $E_T$ distributions in this $E_T$ region (tail). This
phenomenon does not affect
the (true) ratio $J/\psi$ over $DY$ (obtained by the NA50 standard
analysis), but does affect the one
obtained by the so-called minimum bias analysis. A good agreement is
obtained with the data coming
from both analysis -- as well as with the ratios of $J/\psi$ and $DY$
over minimum bias -- in the whole $E_{T}$ region.
\end{abstract}

\vskip 1 truecm

\noi LPT Orsay 01-42 \par
\noi May 2001\par
\newpage
\pagestyle{plain}
\section{Introduction}
\hspace*{\parindent} The NA50 collaboration has observed
\citm{1r}{3r} an anomalous $J/\psi$
suppression in $Pb$ $Pb$ collisions, i.e.  a suppression
significantly larger than the one expected
from the extrapolation of $pA$ and $SU$ data. Furthermore, the shape
of the ratio $J/\psi$ over $DY$
shows a clear change of curvature -- from concave to convex -- at
$E_T \sim 100$~GeV close to the knee
of the $E_T$ distribution followed by a steady
decrease at larger $E_{T}$. \par

It is important for the discussion in this paper to distinguish
between two types of analysis of the
data \cite{4r}. One is the standard analysis, in which the ratio
$J/\psi$ over $DY$ is measured. The
other one is called the minimum-bias analysis. In the latter only the
ratio of $J/\psi$ over
minimum bias ($MB$) is measured -- and is multiplied by a theoretical
ratio $MB$ over $DY$, i.e.

\beq
\label{1e}
\left ( {J/\psi \over DY} \right )_{MB\ analysis} = \left ( {J/\psi
\over MB} \right )_{exp.} \left
( {MB \over DY} \right )_{th}^{NA50} \ . \eeq

\noi Data for the true ratio $J/\psi$ over $DY$, i.e. obtained in the
standard analysis, are only
available up to the knee of the $E_T$ distribution.
They give some indication of the change of curvature at $E_T
\sim \ 100$~GeV mentioned above. However,
the information on the behaviour of this ratio at larger
values of $E_T$ comes entirely from the $MB$ analysis. In this
analysis, the second factor in
(\ref{1e}) is practically constant for $E_T > 100$~GeV and therefore
the behaviour of the ratios
$J/\psi$ over $DY$ and $J/\psi$ over $MB$ in this large $E_T$ region
is practically the same.

In this paper we present a simple
ansatz according to
which the ratio $DY$ over
$MB$ is not constant at large
$E_T$ but decreases for $E_T > 100$~GeV, i.e. its
behaviour is qualitatively
similar to the one of the ratio $J/\psi$ over $MB$. As a consequence,
the true ratio $J/\psi$ over $DY$ for
$E_T > 100$~GeV is significantly flatter than the one obtained according
to eq. (\ref{1e}) -- and in good
agreement with the predictions \citd{5r}{6r} based on
comovers interaction, taking into account
the fluctuations in the
density of comovers at large $E_T$. \par

Our ansatz is based on the trivial observation that, in the $J/\psi$
event sample, $E_{T} \sim
3$~Gev is taken by the $J/\psi$ trigger and, thus, the transverse
energy deposited in the calorimeter
by the other hadron species will be slightly smaller than the
corresponding one in the $MB$ event
sample. This decrease is, of course, very small.
However, its effect on the ratio of $J/\psi$ over $MB$ at the tail of
the $E_T$ distribution is much
larger.
A possible way to determine this hadronic $E_{T}$ loss is the following (see
Section 2c for a more detailed discussion). Assume that one pair of
participants, out of $n_{A}$, produces the $J/\psi$ and no other
hadrons, except, of course, in the fragmentation regions of the
collision of this pair (which will
contain at least the two leading baryons).
The binary collisions of the remaining $n_{A} - 1$ pairs are assumed to
produce ordinary hadrons, exactely in the same way as in the $MB$
event sample. In this case, the  $E_{T}$ loss in the $J/\psi$
event sample at $b \sim 0$ (where $n_{A}$ is close to 200)
is about 0.5~\%. In order to have a rough estimation of
its effect on the $J/\psi$ over $MB$ ratio at the tail of the
$E_{T}$ distribution let
us consider the $E_T - b$
correlation $P(E_T, b) \propto \exp
\left \{ - \left ( E_T - E_T(b) \right )^2/2qa\ E_T(b) \right .$ With
the values of the parameter
given below (Section 2a) we find at $E_{T} = 130$~GeV~:

\beq
\label{2e}
{P_{MB} (E_T = 130 \ {\rm GeV}, b = 0) \over P_{J/\psi} (E_T = 130 \
{\rm GeV}, b = 0)} \simeq {\exp
\left \{ - (130 - E_T(0))^2/2qa \ E_T(0) \right \} \over \exp \left
\{ - (130 - 0.995 E_T(0))^2/2qa\
0.995 E_T(0) \right \}} = 1.3 \ .
\eeq

\noi Such an effect is close to what is needed
in the model of refs. \citd{5r}{6r}
in order to reproduce
the data. This shows
that a small variation in the value of $E_T(0)$ between the $J/\psi$
and the $MB$ event samples
(0.5~\%), produces a much larger effect in their ratio at the tail of the $E_T$
distribution. \par
 
A more precise calculation of the hadronic
$E_{T}$ loss and of its effect on the ratio
$J/\psi$ over $MB$ beyond the knee will be given below. However,
independently of any details, we see that one should be very cautios
in interpreting the experimental results on this ratio
beyond the knee of the $E_{T}$ distribution. \par

Clearly, the effect discussed above will affect the ratio $DY$ over
$MB$ in a similar way -- while the
ratio $J/\psi$ over $DY$ turns out to be practically unaffected. Of course,
the ratio of $J/\psi$ over $DY$,
obtained according to eq. (\ref{1e}), will be affected in the same
way as the ratio $J/\psi$ over
$MB$. \par

The plan of this paper is as follows. In Section 2a we describe a
model for $J/\psi$ suppression
based on comovers interaction with the comovers density computed in
the Dual Parton Model (DPM). We
show that in the rapidity region of the dimuon trigger, the charged
multiplicity per participant
increases with centrality, whereas in that of the $E_T$ calorimeter
it is centrality independent.
In Section 2b we introduce fluctuations in the density of comovers
beyond the knee of the $E_T$
distribution. In Section 2c we propose a simple ansatz to compute the
loss in the average $E_T(b)$
of the $J/\psi$ event sample resulting from the trigger requirement.
In Section 3 we present our
results for the ratios $J/\psi$ over $MB$, $DY$ over $MB$
and $J/\psi$ over $DY$. As
explained above the latter is found to be different in the standard
analysis and in the $MB$ one.
Conclusions are given in Section 4.

\section{The model}
\subsection*{a) Comovers interaction in the dual parton model}
\hspace*{\parindent}
Here we summarize the model of $J/\psi$ suppression based on comovers
interaction \citm{5r}{8r}. We
use the centrality dependence of the comovers
density obtained \cite{10r} in the
Dual Parton Model (DPM) \cite{10newr}. The
model in this subsection can be used from
peripheral collisions up to the knee of the
$E_T$ distribution. To go into the
tail of the distribution one has to introduce the modifications in
Sections 2b and 2c. \par

The cross-section of $MB$, $DY$ and $J/\psi$ event samples are given by
 
\beq
\label{2bise}
I_{MB}^{AB} (b) \propto \sigma_{AB} (b)
\eeq

\beq
\label{3e}
I_{AB}^{DY}(b) \propto \int d^2s \ \sigma_{AB}(b) \ n(b, s)
\eeq

\beq
\label{4e}
I_{AB}^{J/\psi}(b) \propto \int d^2s \ \sigma_{AB}(b) \ n(b,s) \
S_{abs}(b, s) S_{co}(b, s) \quad .
\eeq

\noi Here $\sigma_{AB} (b) = \int d^2s \{ 1 - \exp [ - \sigma_{pp} \
AB\ T_{AB}(b, s)] \}$ where
$T_{AB}(b, s) = T_A (b) \ T_B(b - s)$ is the product of profile
functions obtained from the
Woods-Saxon nuclear densities \cite{9r}. Upon integration over $b$ we
obtain the $AB$ total
cross-section, $\sigma_{AB}$. The factor $n$ in (\ref{2bise}) is given by

\beq
\label{5e}
n(b, s) = AB \ \sigma_{pp} \ T_A(s) \ T_B(b - s)/\sigma_{AB}(b) \quad .
\eeq

\noi Upon integration over $s$ we obtain the average number of binary
collisions $n(b) = AB \
\sigma_{pp} \ T_{AB}(b)/\sigma_{AB}(b)$. \par

The factors $S_{abs}$ and $S_{co}$ in (\ref{3e}) are the survival
probabilities of the $J/\psi$
respectively due to nuclear absorption and comovers interaction. They
are given by

\beq
\label{6e}
S^{abs}(b, s) = {[1 - \exp (- AT_A(s) \ \sigma_{abs})] [1 - \exp (- B
\ T_B (b - s) \ \sigma_{abs})]
\over \sigma_{abs}^2 \ AB \ T_A(s) \ T_B(b - s)} \eeq

\beq
\label{7e}
S^{co} (b, s) = \exp \left [ - \sigma_{co} {3 \over 2}
N_{y_{DT}}^{co}(b, s) \ell n \left ( {{3
\over 2} N_{y_{DT}}^{co}(b, s) \over N_f} \right ) \right ]  \eeq

\noi In (\ref{7e}), $N_{y_{DT}}^{co}(b, s)$ is the density of charged
comovers (positives and
negatives) in the rapidity region of the dimuon trigger and $N_f =
1.15$~fm$^{-2}$ \citd{7r}{8r}
is the corresponding density in $pp$. The factor 3/2 in (\ref{7e})
takes care of the neutrals.
In the numerical calculations we use
$\sigma_{abs} = 4.5$~mb \footnote{This value is in between the ones
obtained from fits of the $pA$ data of the NA38 \cite{18r} and
of E866 \cite{19r} Collaborations, and is consistent with both
sets of data.} and
$\sigma_{co} = 1$~mb \citd{5r}{6r}.
\par

In order to compute the density of comovers we use the DPM formalism
developed in \cite{10r}. The
density of charged particles is given by a linear superposition of the
density of participants and
the density of binary collisions with coefficients calculable in DPM.
For $A = B$ we have

\bea
\label{8e}
&&N_y^{co}(b) = n_A(b) \left [ N_{\mu (b)}^{qq^P-q_{v}^T}(y) +
N_{\mu (b)}^{q_{v}^P-qq^T}(y) +
(2k - 2) \ N_{\mu (b)}^{q_s-\bar{q}_s} \right ] \nonumber \\
&&+ [(n(b) - n_A(b)] \ 2k \ N_{\mu(b)}^{q_s-\bar{q}_s} (y) \quad . \eea

\noi Here $n_A(b, s)$ is given by
\beq
\label{9e}
n_A(b, s) = A \ T_A(s) \left [ 1 - \exp \left \{ - \sigma_{pp} \ B \
T_B (b - s) \right \} \right
]/\sigma_{AB}(b) \quad . \eeq

\noi Upon integration over $s$ of (\ref{9e}) we obtain the average
number of participants of $A$ (or
participant pairs for $A = B$). $k = 1.4$ \cite{10r}
is the average number of
inelastic collisions in each $NN$
collision and $\mu (b) = k \ n(b)/n_A(b)$ is the total average number
of collisions suffered by each
participant. The first term in (\ref{8e}) is the charged plateau
height in one $NN$ collision,
resulting from the superposition of $2k$ strings, multiplied by the
average number of participant
pairs $n_A$. This would be the only contribution if each participant
of $A$ would interact with
only one participant of $B$. The second term in (\ref{8e}) contains
the contributions resulting
from the extra collisions of each participant. Since in DPM there
are two strings per inelastic
collision, this second term, consisting of strings stretched between
sea quarks and antiquarks,
completes the total average number of strings -- equal to $2kn$. The
string multiplicities in
(\ref{8e}) are obtained from a convolution of momentum distribution
functions and fragmentation
functions. All relevant equations and the values of the parameters
needed for their calculation are
given in \cite{10r}. Their numerical values for several values of $b$
are listed in Table 1 -- for the
rapidity regions of both the dimuon trigger and the $E_T$ calorimeter. \par

The results, in the rapidity regions of
both the dimuon trigger and
the $E_T$ calorimeter, are shown in Fig.~1. We see that the charged
multiplicity per participant
increases with centrality in the rapidity region of the dimuon
trigger. This increase is slightly
smaller than the one in the central region $- 0.5 < y^* <
0.5$ \citd{10r}{11r}.
However, it is still  significant. On the contrary, in the rapidity
region of the calorimeter,
the same quantity is practically independent of centrality and we
recover here the behaviour
expected in the wounded nucleon model \cite{12r}. \par

In order to compute $S^{co}(b, s)$, eq. (\ref{7e}), we need the 
density of comovers at each
$b$ and $s$. In this case $n_A \not= n_B$ and we have to use the 
general DPM formulae (eqs.
6.1 and 6.15 of \cite{10newr}). We have \cite{8r}

\bea \label{11enew}
&&N_y^{co}(b,s) = \left [N_1 \ n_A(b,s) + N_2 \ n_B(b, b-s) + N_3 \ 
n(b,s)\right ] \theta \left ( n_B (b, b-s) - n_A(b, s) \right ) \nn \\
&&+ \left [ N'_1\  n_A(b,s) + N'_2 \ n_B (b, b-s) + N'_3 \ n(b, 
s)\right ] \theta \left ( n_A(b,s) - n_B (b, b -s) \right ) \ . \eea

\noi The values of the coefficients $N_i$ and $N'_i$ are given in 
Table 2 \footnote{Note that $N_i$ and $N'_i$ depend on
centrality via $\mu (b)$ (see eq. (\ref{8e})). This dependence has 
been neglected in \cite{8r} -- where an average value of
$\mu$ was used.}. 
By comparing the values in Tables 1 and 2 we see that 
for $n_A = n_B$ we recover eq. (\ref{8e}). \par

Eqs. (\ref{2e}) to (\ref{9e}) allow to compute the impact parameter 
distributions of the
$MB$, $DY$ and $J/\psi$ event samples. Experimental results for these
quantities are plotted as a
function of observable quantities such as $E_T$~--
the energy of neutrals deposited in the calorimer. Using the
proportionality between $E_T$ and
multiplicity, we have

\beq
\label{10e}
E_T(b) = {1 \over 2} \ q \ N_{y_{cal}}^{co}(b) \quad .
\eeq

\noi Here the multiplicity of comovers is determined in the rapidity
region of the $E_T$
calorimeter. The factor $1/2$ is introduced because $N^{co}$ is the
charged multiplicity
whereas $E_T$ refers to neutrals. In this way $q$ is close to the
average transverse energy per
particle, but it also depends on the calibration of the calorimeter.
The correlation $E_T - b$
is parametrized in the form

\beq
\label{11e}
P (E_T, b) = {1 \over \sqrt{2 \pi q a E_T(b)}}
\ \exp \left \{ - (E_T - E_T(b))^2/2 q a E_T(b)
\right \} \quad .
\eeq

\noi The $E_T$ distributions of $MB$, $DY$ and $J/\psi$ are then
obtained by folding eqs.
(\ref{2bise})-(\ref{4e}) with $P(E_T, b)$, i.e.

\beq
\label{12e}
I_{AB}^{MB,DY,J/\psi}(E_T) = \int d^2b \ I_{AB}^{MB,DY,J/\psi}(b) \
P(E_T, b) \quad . \eeq

\noi The parameters $q$ and $a$ are obtained from a fit of the $E_T$
distribution of the $MB$ event
sample. Note that since
$N_{y_{cal}}^{co}(b)$ is proportional to the number of participants
(see Fig. 1) our fit is
identical to the one obtained \cite{7r} using the wounded nucleon model.
Actually, we obtain identical curves to the ones
in Fig. 1 of ref. \cite{1r} -- where the $E_T$ distributions of $MB$
events of 1996 and 1998 are
compared with each other. The values of the parameters for the 1996
data are $q = 0.62$~GeV and $a =
0.825$. For the 1998 data, the tail of the $E_T$ distribution is
steeper, and we get $q = 0.62$~GeV
and $a = 0.60$\footnote{Note that the same
value of the parameter $a$ is used in the $MB$, $DY$ and $J/\psi$
event sample. A priori there could
be some differences in the fluctuations for hard and soft processes.}.
In the following we shall use the latter values. Indeed, according
to the NA50 Collaboration \cite{4r}, the 1996 data (thick target)
at large $E_T$ are contaminated by rescattering effects -- and only
the 1998 data should be used beyond the knee.

\subsection*{b) Comovers fluctuations}
\hspace*{\parindent} The model described above allows to compute the
$E_T$ distribution of $MB$,
$DY$ and $J/\psi$ event samples between peripheral $AB$ collisions
and the knee of the $E_T$
distribution. Beyond it, most models, based on either deconfinement or
comovers interaction, give a
ratio $J/\psi$ over $DY$ which is practically constant -- in
disagreement with NA50 data. A possible
way out was suggested in \cite{5r}. The idea is that, since $E_T$
increases beyond the knee due to
fluctuations, one can expect that this is also the case for the
density of comovers. Since
$N_{y_{DT}}^{co}$ does not contain this fluctuation it has been
proposed to introduce the
following replacement in eq. (\ref{7e})~:

\beq
\label{13e}
N_{y_{DT}}^{co}(b, s) \to N_{y_{DT}}^{Fco}(b, s) = N_{y_{DT}}^{co}(b, s) \ F(b)
\eeq

\noi where $F(b) = E_T/E_T(b)$. Here $E_T$ is the measured value of
the transverse energy and
$E_T(b)$ is its average value given by eq. (\ref{10e}) -- which does
not contain the fluctuations.
\par

The replacement (\ref{13e}) assumes that the fluctuation in
$N_y^{co}$, in the rapidity region of
the dimuon trigger, is identical to the fluctuation in $E_T$ measured
by the calorimeter. Since
these two regions do not overlap, the two
fluctuations could be weakly correlated. However, this is not the
case in string models where multiplicity fluctuations
are mainly due to fluctuation in the number of
strings (rather than to multiplicity fluctuations within individual
strings). This introduces
long-range rapidity correlations \cite{15r} (see also \cite{16r} for
a discussion on this point).
\par

It has been shown in \cite{14r} that, introducing the fluctuations
given by (\ref{13e}) in a
deconfining approach, one can describe the NA50 data, at large and
intermediate values of $E_T$,
either with two sharp deconfining tresholds or with a gradual onset of $J/\psi$
suppression (rather than a sharp one)\footnote{Note, however, that this model
does not describe the data at low $E_T$. Moreover, according to the
anaysis of \cite{6r}, it
should also fail to describe the preliminary data \cite{2r} on the
$J/\psi$ suppression versus the
energy $E_{ZDC}$ of the zero degree calorimeter for peripheral
collisions. Indeed, it was
shown in \cite{6r} that these data confirm the low $E_T$ shape of the data
versus $E_T$. Moreover,
this new data require the anomalous suppression
to be present for very
peripheral $Pb$ $Pb$ collisions.}. \par

Introducing the replacement (\ref{13e}) in the comovers model of
Section 2a, we obtain \cite{5r} a change of
curvature in the ratio $J/\psi$ over $DY$ at $E_T \sim 100$~GeV. However, its
decrease at larger values of $E_T$ is smaller than the experimental
one. In the next section, we
introduce a new mechanism which increases the $J/\psi$ suppression at
large $E_T$, in the $MB$ analysis.

\subsection*{c) $E_T$ loss of ordinary hadrons induced by the $J/\psi$ trigger}
\hspace*{\parindent} As discussed in the Introduction,
due to energy conservation, the production
of a $J/\psi$ in the dimuon
trigger produces a decrease of the $E_T$ of ordinary hadrons in the
$J/\psi$ event sample -- as
compared to its value in the $MB$ one. As discussed there, this
effect is strongly amplified in the
tail of the $E_T$ distribution. The size of this
hadronic $E_{T}$ loss is very small. However,
its exact value is difficult to compute since we do not
know in detail the rapidity
region affected by the about 3 GeV of $E_T$ taken by the $J/\psi$. We
propose the following ansatz. Let us
single out the $NN$ collision in which the $J/\psi$ is produced and
let us assume that in this
collision no hadrons, other than the $J/\psi$, are produced in the
central rapidity interval $- 1.8 \ \lsim
\ y^* \lsim \ 1.8$ -- which includes the regions of the $E_T$
calorimeter and of the dimuon
trigger. Of course, ordinary hadrons, in particular the two leading
baryons, will be produced in
the two fragmentation regions of this collision. With this
assumption, eq. (\ref{10e}) has to be
replaced by

\bea
\label{14e}
&&E_T(b) = {1 \over 2} \ q \left \{  \left ( n_A(b) - 1 \right ) \left [N_{\mu
(b)}^{qq^P-q_{v}^T}(y) + n_{\mu (b)}^{q_{v}^P-qq^T}(y) + (2 k - 2)
N_{\mu (b)}^{q_s-\bar{q}_s}(y)
\right ] \right . \nn \\ &&\left . + \left ( n(b) -
n_A(b) \right ) 2k \ N_{\mu (b)}^{q_s-\bar{q}_s}(y) \right \}  \eea

One could think that the hadronic $E_T$ loss resulting from the replacement of
eq. (\ref{10e}) by eq.
(\ref{14e}) is the maximal possible one. Actually, this is not the
case. Indeed, it could happen
that the $J/\psi$ trigger produces a decrease of the average energy
of the other collisions
involving either one of the two nucleons of the binary collision
which produces the $J/\psi$. In (\ref{14e}) we
have assumed that this is not the case and that all these extra
collisions produce exactly the
same hadronic multiplicity as in the $MB$ case.
One could  instead assume that
no ordinary hadrons are produced in these extra collisions either,
i.e. that one pair of participants
produces only the $J/\psi$ and no other hadrons (in the central
rapidity interval defined above).
In this case (\ref{10e}) should be scaled down by a factor
$(n_A(b) - 1)/n_A(b)$, i.e.

\beq
\label{15e}
E_T(b) = {1 \over 2} \ q \ {n_A(b) - 1 \over n_A(b)} \
N_{y_{ca}}^{co}(b) \quad .
\eeq

\noi Both possibilities, eqs. (\ref{14e}) and (\ref{15e}), will be
discussed in the next
section\footnote{The same changes
should also be made in
$N_{y_{DT}}^{co}$ but, obviously, their effect is negligibly small.}.
Of course, for the $MB$ event sample we use eq. (\ref{10e}) without
any change. \par

In order to see how the hadronic $E_T$ loss in the $J/\psi$ event sample,
given by eqs. (\ref{14e}) and (\ref{15e}), compares to the
$E_T \sim 3$~GeV taken by the $J/\psi$, we have calculated the charged
multiplicity per participant pair in the rapidity region
$-1.8 < y^{*} < 1.8$ -- which includes both the dimuon trigger and
the $E_{T}$ calorimeter.
Using the values in Table 1 we obtain 6.6. Multiplying it by 1.5 to
include the neutrals and assuming an average $E_{T}$ per particle of
0.5 GeV we obtain an average $E_{T}$ loss of $E_{T} \sim 5$~GeV.
This is the average $E_{T}$ loss induced by eq. (\ref{15e}), where
the contribution of one participant pair
to the hadronic $E_{T}$ has been
removed. At $b = 0$, where $n_{A}$ is close to 200, it amounts to
about 0.5~\% of the total $E_{T}$ produced in the above rapidity interval.
From, the values in Table 1, one can also see that the average $E_{T}$ loss
induced by (\ref{14e}) is about one half of the above value, i. e.
$E_{T} \sim 2.5$~GeV -- corresponding to about 0.25~\% of the total
$E_{T}$ in the considered rapidity interval. Therefore an average
$E_{T}$ loss of $E_{T} \sim 3$~GeV would be somewhere in between
the ones obtained with the ansatzs in eqs. (\ref{14e}) and (\ref{15e}).

\section{Numerical results and comparison with\break \noindent experiment}

\subsection*{a) Ratio $J/\psi$ over $MB$}
\hspace*{\parindent} The results for the ratio $J/\psi$ over $MB$
versus $E_T$ in $Pb$ $Pb$
collisions at $158$~GeV are shown in Fig.~2 and compared
with NA50 preliminary data
\cite{2r}\footnote{The points obtained with the standard analysis
are not included in Fig.~2 since they are not experimental
measurements of the ratio $J/\psi$ over $MB$. These points extend
to lower values of $E_T$ than the ones obtained in the $MB$ analysis.
Unfortunately, we see from Fig.~7 of \cite{4r} that, at low $E_T$,
the measured values of $DY$ over $MB$ deviate from the theoretical ones.}.
The dotted curve is the result of the model as presented
in Section 2a, i.e. without
the effect of the fluctuations, eq. (\ref{13e}), or that of the $E_T$
loss induced by eqs.
(\ref{14e}) or (\ref{15e}). The dashed line is the result obtained
with the fluctuations but without
the $E_T$ loss. The dashed-dotted and the solid curves are
the results obtained when both
effects are taken into account using eqs. (\ref{14e}) and (\ref{15e}),
respectively, for the $E_{T}$ loss.
We see that the turn-over and subsequent decrease are well
reproduced. Note that the calculation of the dashed,
dashed-dotted and full lines
does not involve any new free
parameter. \par

\subsection*{b) Ratio $DY$ over $MB$}
\hspace*{\parindent} As mentioned in Section 2c the effect of the
$E_T$ loss should also be present
in the case of the $DY$ event sample -- where a dimuon of average
mass close to that of $J/\psi$ is
produced. We use in this case the same prescriptions, eqs.
(\ref{14e})-(\ref{15e}), as for
$J/\psi$. The results for this ratio versus $E_T$ in $Pb$ $Pb$
collisions at $158$~GeV
are presented in Fig.~3 and compared to 1996 NA50 data \cite{4r}.
The dotted line is obtained
without introducing the $E_T$
loss. We see that the ratio tends to saturates at large $E_T$.
The dashed-dotted and solid
curves are
obtained using the $E_T$ loss given by eqs. (\ref{14e}) and
(\ref{15e}), respectively. We observe a turn-over
and a subsequent decrease at large $E_T$, qualitatively similar to
the one in $J/\psi$ over $MB$
ratio, but smaller in magnitude due to the absence, here, of the
comover interaction -- and, hence, of the effect of the
fluctuations. It is
interesting that the data, although having large statistical errors,
also seem to indicate a
turn-over. In any case, they are consistent with our
results\footnote{Note that the ration $DY$ over $MB$ has been
measured in 1996. The NA50 Collaboration does not consider the
1996 data to be reliable beyond the knee due to possible
contamination of rescattering in the (thick) target.
However, from the comparison of 1996 and 1998 data on
the ratio $J/\psi$ over $MB$,
we can expect that, using a thin target, the $DY$ over
$MB$ ratio beyond the knee would be smaller than
with a thick one.}. \par

A test of our mechanism could be provided by a measurement of the
ratio open charm over $MB$
beyond the knee.
The presence of a similar turn-over beyond the knee would
strongly support our interpretation.

\subsection*{c) Ratio $J/\psi$ over $DY$}
\hspace*{\parindent} The results of our model for this ratio, versus
$E_T$, in $Pb$ $Pb$ collisions at
$\sqrt{s} = 158$~GeV, are shown in Fig.~4 and compared to NA50 data
\citm{1r}{3r} -- both for the
true $J/\psi$ over $DY$ ratio (labeled with $DY$), and for the one
obtained with the $MB$ analysis
(labeled Min. Bias). The results of the model for the true ratio
are given by
the dotted line (without fluctuations) and
the dashed line (with fluctuations). The latter is
practically the same as
the ones obtained in
\cite{5r} and \cite{6r}. In both cases the $E_T$ loss mechanism
introduced above was not present.
However, our results for the true ratio $J/\psi$ over $DY$ do not
change, since the effect due to
the $E_T$ loss cancels, to a high accuracy, in this ratio.
We see that our results are in good agreement with the NA50
data for this ratio -- which
do not extend beyond the knee.
The other data in
Fig.~4 are obtained with the $MB$ analysis, summarized by eq. (\ref{1e}) --
where the last factor is given
by eq.~(\ref{12e}), and, therefore, does not contain the $E_T$ loss
induced by eqs. (\ref{14e}) or (\ref{15e}). Using our results
(Section 3a) for the ratio $J/\psi$
over $MB$ and multiplying it by the ratio $(MB/DY)_{th}^{NA50}$ we
obtain the dashed-dotted and solid curves in Fig.~4.
The agreement with the NA50 data obtained with the $MB$
analysis is substantially improved. \par

A straightforward consequence of the above results is that the data
of the standard and $MB$
analysis
are different at large $E_{T}$ and, therefore, should not
be put on the same figure. The latter should be shown
only as a ratio $J/\psi$ over
$MB$ (which is the measured quantity in this analysis). A plot of
the $J/\psi$ over $DY$ ratio
should include only data obtained in the standard analysis.

Another consequence of our approach is that the decrease, beyond the knee,
of the ratio $J/\psi$ over $DY$ in the $MB$ analysis, as a function
of $E_{ZDC}$ (the energy of the zero degree calorimeter),
should be less pronounced than the corresponding one versus $E_{T}$.
This is due to the fact that
the main contribution to $E_{ZDC}$ is proportional to the
number of spectators, which is not affected by
the presence of the $J/\psi$ trigger.
The data \citd{2r}{3r} seem to indicate that
this is indeed the case \cite{6r}.

\section{Conclusions}
\hspace*{\parindent} The idea of using the $J/\psi$ over $MB$ ratio
as a measure of the $J/\psi$
suppression was first introduced in ref. \cite{17r}. While this
allows to improve considerably
the statistics, it can only be safely used from peripheral collisions
up to the knee of the
$E_T$ distribution. Beyond it, this ratio is very sensitive to small
differences between the
average $E_T$ of the $J/\psi$ and $MB$ event samples. We have
argued that such a difference is
indeed present due to the $E_T$ taken by the $J/\psi$ trigger. \par

We have introduced a simple, physically sound, ansatz to evaluate
this $E_T$ loss in the
$J/\psi$ event sample (as compared to its value in the $MB$ one) and
have shown that with this
ansatz the NA50 data for the ratio $J/\psi$ over $MB$ can be
reproduced. In our model, this
mechanism does not affect the true ratio $J/\psi$ over $DY$ -- i.e.
the one obtained by the
standard analysis. However, it does affect this ratio in the case of
the minimum bias
analysis. Agreement is obtained with the NA50 data for this ratio, 
in both analysis. \par

An unavoidable consequence of our approach is that the ratio $DY$
over $MB$ should also exhibit
a turn-over close to the knee of the $E_T$ distribution followed by a
subsequent decrease. Actually,
the 1996 NA50 data \cite{4r} provide some indication of such a turn-over.
Our results are in
agreement with these data. \par

Independently of the details of our
implementation of the hadronic $E_T$ loss
in the $J/\psi$ event sample, the analysis presented here indicates
that it is premature to
interpret the NA50 data at large $E_T$ (beyond the knee of the $E_T$
distribution) as a
manifestation of a phase transition to a new state of matter.
Actually, we have shown that a
model based on comovers interaction provides a good description of
the NA50 data in the
whole $E_T$ region. \\

\noi {\large \bf Acknowledgments} \\

It is a pleasure to thank N. Armesto, E. G. Ferreiro, D. Kharzeev,
C. Salgado and J. Tran Thanh Van for discussions.
D.S. thanks Fundaci\'on Barrie de la Maza for
financial support. This work was supported in part
by NATO grant PSTCLG 977275. \\

\def\labelenumi{[\arabic{enumi}]}
\noindent
{\large\bf References}

\ben
\item\label{1r} NA50 collaboration, M. C. Abreu et al., Phys. Lett.
{\bf B477}, 28 (2000).
 
\item\label{2r} NA50 collaboration, Proceedings QM 2001, presented by
P. Bordalo (to be published
in Nucl. Phys. A).
 
\item\label{3r} NA50 collaboration, Proceedings XXXVI Rencontres de
Moriond, Les Arcs, France 2001, presented by R. Arnaldi.

\item\label{4r} NA50 collaboration, Phys. lett. {\bf B450}, 2456 (1999).

\item\label{5r} A. Capella, E. G. Ferreiro and A. Kaidalov, Phys.
Rev. Lett. {\bf 85}, 2080 (2000).

\item\label{6r} A. Capella and D. Sousa, nucl-th/0110072. 
N. Armesto, A. Capella, E. G. Ferreiro, A. Kaidalov
and D. Sousa, nucl-th/0104004, Proceedings QM
2001, presented by A. Capella, ibid, and Proceedings XXXVI Rencontres de
Moriond, Les Arcs, France 2001, presented by D. Sousa.

\item\label{7r} D. Kharzeev, C. Louren\c co, M. Nardi and H. Satz, Z.
Phys. {\bf C74}, 307 (1997).

\item\label{8r} N. Armesto and A. Capella, Phys. Lett. {\bf B430},
23 (1998). N. Armesto, A.
Capella and E. G. Ferreiro, Phys. Rev. {\bf C59}, 359 (1999).

\item\label{10r} A. Capella and D. Sousa, Orsay preprint LPT
00-137, nucl-th/0101023, to be published in Phys. Lett. {\bf B}.

\item\label{10newr} A. Capella, U. Sukhatme, C-I Tan and J. Tran 
Thanh Van, Phys. Rep. {\bf 236}, 225 (1994).

\item\label{9r} C. W. de Jager et al, Atomic Data and Nuclear Data
Tables 14, 485 (1974).

\item\label{11r} WA98 collaboration, M. M. Agarwal et al., nucl-ex/0008004.

\item\label{12r} A. Bialas, A. Bleszyncki and W. Cyz, Nucl. Phys.
{\bf B111}, 461 (1976).

\item\label{15r} A. Capella and A. Krzywicki, Phys. Rev. {\bf D18},
4120 (1978).

\item\label{16r} J. H\"ufner, B. Kopeliovich and A. Polleri, nucl-th/0012003.

\item\label{14r} J. P. Blaizot, P. M. Dinh and J. Y. Ollitrault,
Phys. Rev. Lett. {\bf 85}, 4020
(2000) and Proceedings QM 2001, presented by P. M. Dinh, ibid.

\item\label{17r} A. Capella, J. Ranft, J. Tran Thanh Van, C. Merino,
C. Pajares and A. V. Ramallo, Proceedings 25th International
Conference on High Energy Physics, Singapore 1990 (Editors
K. K. Phua and Y. Yamaguchi). See also J. Gosset, A. Baldisseri,
H. Borel, F. Staley and Y. Terrien, Euro. Phys. Jour. {\bf C13},
63 (2000).

\item\label{18r} NA38 Collaboration, M. C. Abreu et al.,
Phys. Lett. {\bf B406}, 408 (1999).

\item\label{19r} E866 Collaboration, M. J. Leitch et al.,
Phys. Rev. Lett. {\bf84}, 3256 (2000).

\een

\newpage
\centerline{\large \bf Figure Captions :}
\vspace{0.3cm}

\noi {\bf Figure 1.} Charged multiplicity per
unit-rapidity and per
participant pair
versus number of participants in $Pb$ $Pb$ collisions at $158$~GeV,
in the rapidity regions of the dimuon trigger and of the $E_{T}$
calorimeter. \par \vskip 3 truemm

\noi {\bf Figure 2.} The ratio $J/\psi$ over $MB$ versus $E_{T}$
in $Pb$ $Pb$ collisions at $158$~GeV. The dotted curve is obtained
without $E_{T}$ fluctuations and $E_{T}$ loss. The dashed line
contains the $E_{T}$ fluctuations (eq. (\ref{13e})) and no
$E_{T}$ loss. The dashed-dotted and solid lines are obtained when
both effects are taken into account, using, respectively, eqs.
(\ref{14e}) and (\ref{15e}) for the $E_{T}$ loss. The data
(in arbitrary units) are from refs. \citd{2r}{3r}. \par \vskip 3 truemm

\noi {\bf Figure 3.} Same as Fig.~2 for the ratio $DY$ over $MB$.
The dotted  curve is obtained without $E_{T}$ loss.
The dashed-dotted and solid curves are obtained
with $E_{T}$ loss using eqs. (\ref{14e}) and (\ref{15e}),
respectively. The data are from ref. \cite{4r}. \par \vskip 3 truemm

\noi {\bf Figure 4.} Same as Fig.~2 for the ratio $J/\psi$
over $DY$. The data are from ref. \cite{1r}.
The data labeled with $DY$ are for the true $J/\psi$ over $DY$
ratio (standard analysis). They should be compared with the
dotted and dashed lines
obtained, respectively, without and with $E_{T}$ fluctuations. The
effect of the $E_{T}$ loss cancels in the true ratio
$J/\psi$ over $DY$.
The data labeled Min. Bias are obtained with the $MB$ analysis (see
eq. (\ref{1e})) and should be compared with the dashed-dotted and
solid lines, obtained with the $E_{T}$ loss given by eqs.
(\ref{14e}) and (\ref{15e}), respectively. \\

\centerline{\large \bf Table Captions :}
\vspace{0.3cm}

\noi {\bf Table 1.} Values of $N^{qq^{P}-q^{T}_{v}}_{\mu (b)} +
N^{q^{P}_{v}-qq^{T}}_{\mu (b)}$ (second column)
and $N^{q_{s}-\bar{q_{s}}}_{\mu (b)}$ (third column)
for charged particles in
eq. (\ref{9r}), for various values of the impact parameter
($b$) (first column), integrated in $y$ in the rapidity
region of the dimuon trigger $0 < y^{*} < 1$. The forth
and fifth columns are the same quantities integrated in the rapidity
region of the $E_{T}$ calorimeter ($-1.8 < y^{*} < -0.6$),
and divided by its length (1.2 units). \par \vskip 3 truemm

\noi {\bf Table 2.} Values of the coefficients $N_i$ and $N'_i$ in 
eq. (\ref{11enew}), integrated in rapidity in the region
$0 < y^* < 1$ for various values of the impact parameter $b$.

\newpage

\centerline{\bf Fig. 1}
\begin{figure}[hbtp]
\begin{center}
\mbox{\epsfig{file=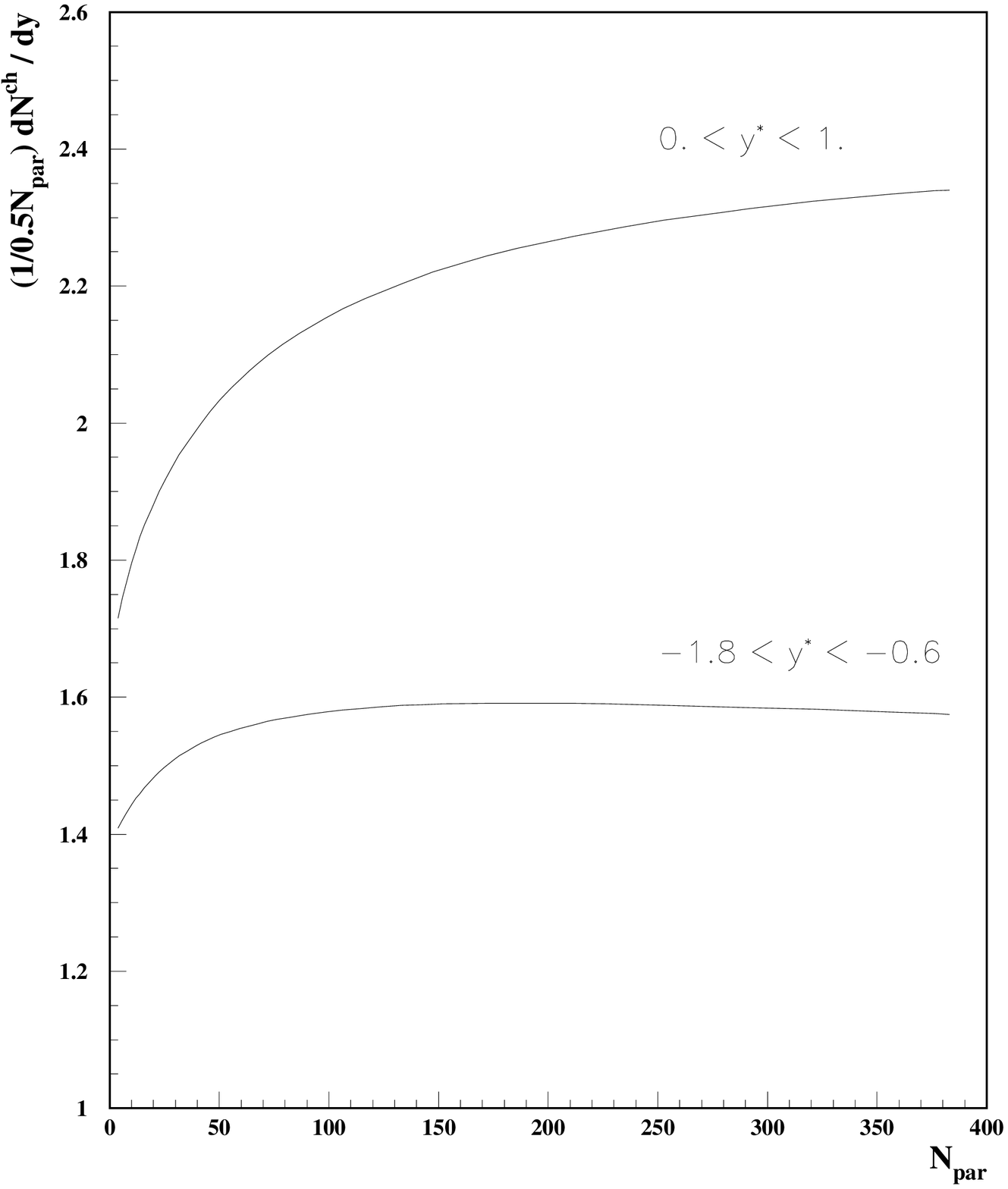,height=16cm}}
\end{center}
\end{figure}

\newpage
\centerline{\bf Fig. 2}
\begin{figure}[hbtp]
\begin{center}
\mbox{\epsfig{file=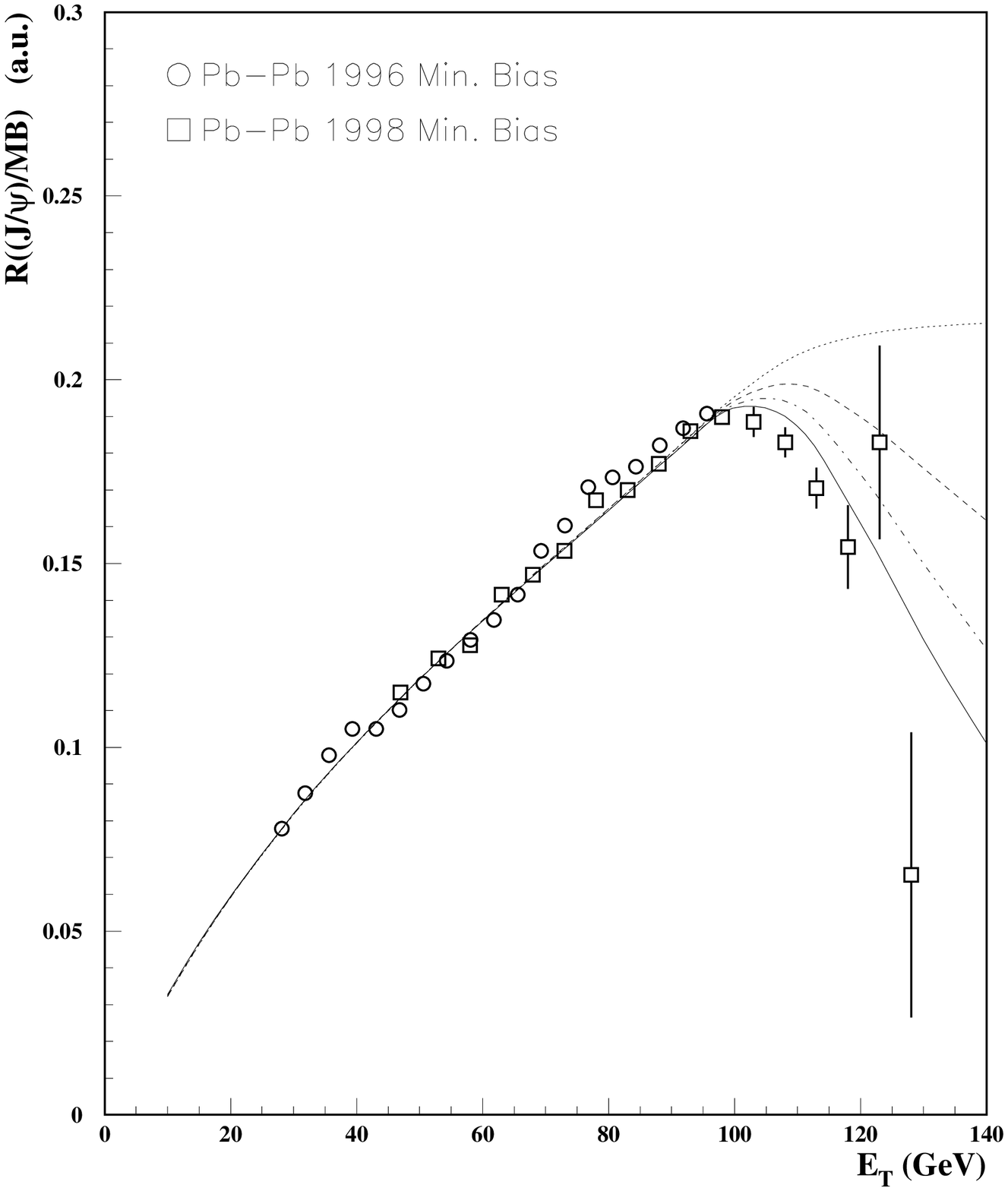,height=16cm}}
\end{center}
\end{figure}

\newpage
\centerline{\bf Fig. 3}
\begin{figure}[hbtp]
\begin{center}
\mbox{\epsfig{file=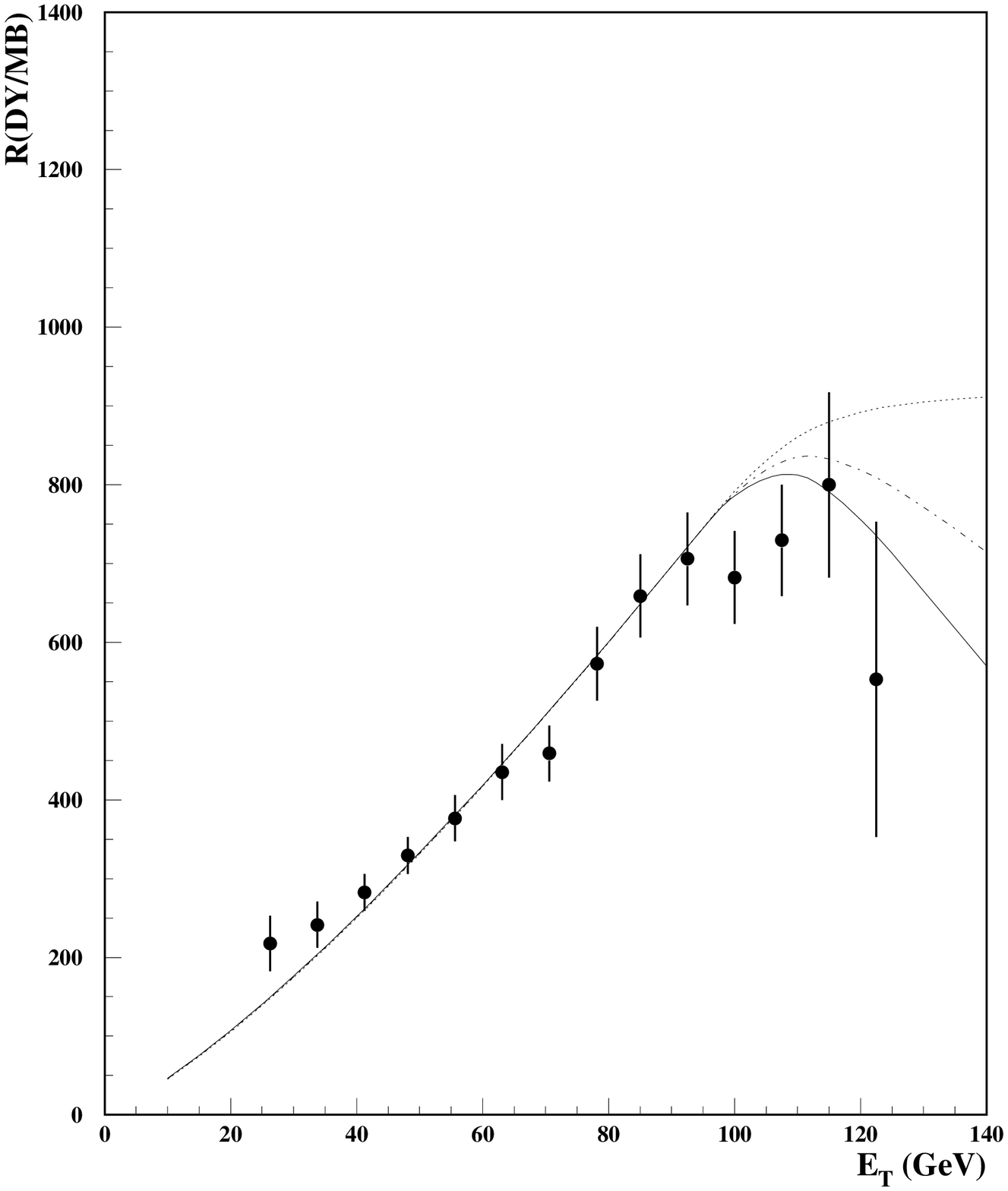,height=16cm}}
\end{center}
\end{figure}

\newpage
\centerline{\bf Fig. 4}
\begin{figure}[hbtp]
\begin{center}
\mbox{\epsfig{file=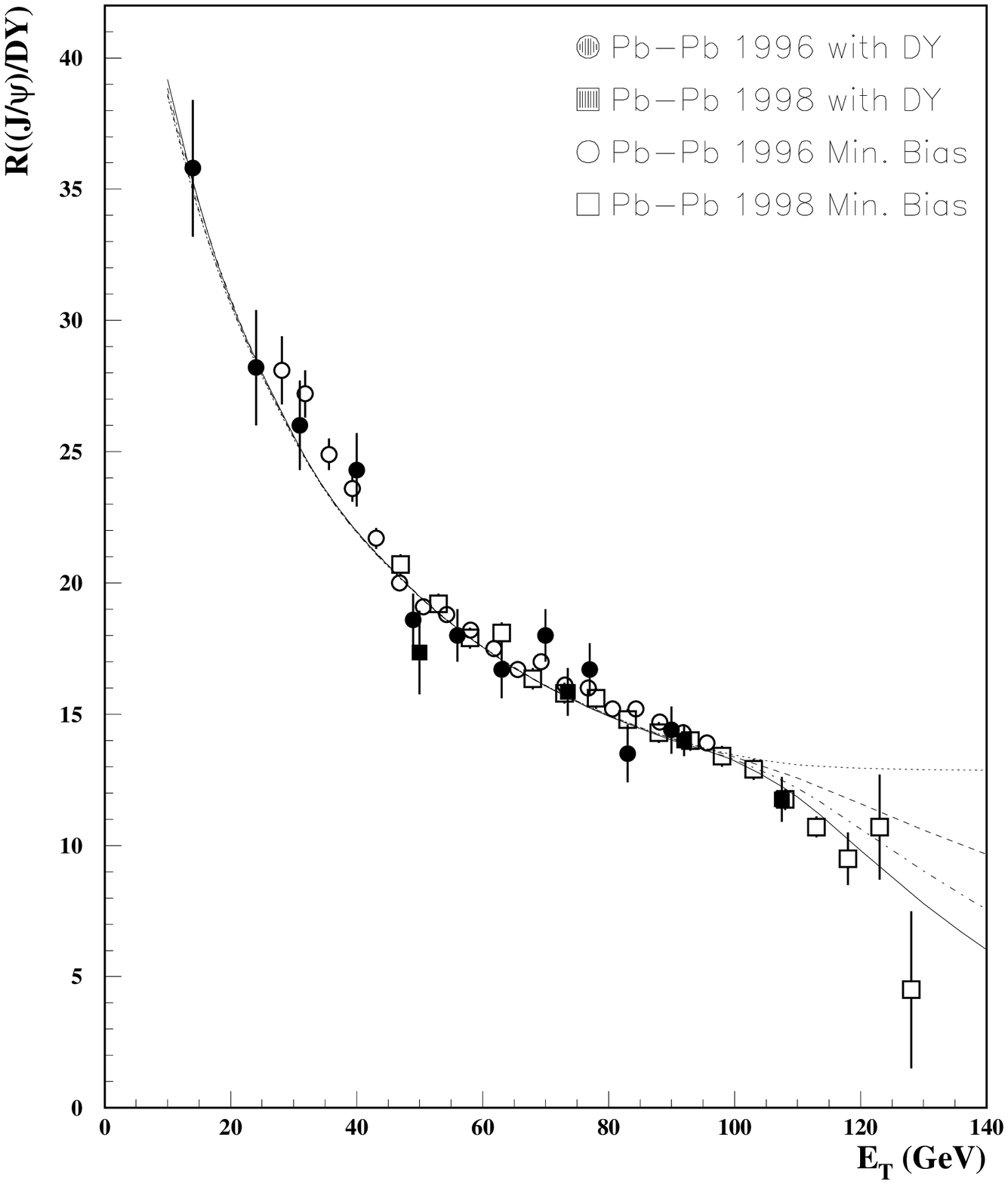,height=16cm}}
\end{center}
\end{figure}

\newpage
\begin{center}
\small{
\begin{tabular}{|c|c|c|c|c|}
\hline
& & & & \\
\qquad $b$ (fm) \qquad & \qquad $N^{qq^{P}-q^{T}_{v}}_{\mu (b)} +
N^{q^{P}_{v}-qq^{T}}_{\mu (b)}$ \qquad &
\qquad $N^{q_{s}-\bar{q_{s}}}_{\mu (b)}$ \qquad &
\qquad $N^{qq^{P}-q^{T}_{v}}_{\mu (b)} +
N^{q^{P}_{v}-qq^{T}}_{\mu (b)}$ \qquad &
\qquad $N^{q_{s}-\bar{q_{s}}}_{\mu (b)}$ \qquad \\
& & & & \\
\hline
& & & & \\
0 & 0.994 & 0.142 & 0.751 & 0.087 \\
2 & 1.002 & 0.144 & 0.760 & 0.088 \\
4 & 1.024 & 0.148 & 0.784 & 0.092 \\
6 & 1.057 & 0.154 & 0.823 & 0.097 \\
8 & 1.105 & 0.164 & 0.878 & 0.106 \\
10 & 1.167 & 0.178 & 0.952 & 0.118 \\
12 & 1.241 & 0.195 & 1.040 & 0.134 \\
& & & & \\
\hline
\end{tabular}
}
\end{center}
\centerline{\bf Table 1}

\vskip 2 truecm

\begin{center}
\small{
\begin{tabular}{|c|c|c|c|c|c|}
\hline
& & & & &\\
\qquad $b$ (fm) \qquad & \qquad $N_1$ \qquad &
\qquad $N_2$ \qquad & \qquad $N_3 = N'_3$ \qquad &
\qquad $N'_1$ \qquad & \qquad $N'_2$ \qquad\\
& & & & &\\
\hline
& & & & & \\
0 & 0.5896 & 0.1210 & 0.3970 & 0.3743 & 0.3363\\
2 & 0.5937 & 0.1222 & 0.4018 & 0.3772 & 0.3387\\
4 & 0.6038 & 0.1250 & 0.4134 & 0.3844 & 0.3444\\
6 & 0.6192 & 0.1295 & 0.4323 & 0.3954 & 0.3533\\
8 & 0.6405 & 0.1358 & 0.4601 & 0.4107 & 0.3656\\
10 & 0.6671 & 0.1443 & 0.4988 & 0.4300 & 0.3814 \\
12 & 0.6961 & 0.1541 & 0.5470 & 0.4513 & 0.3989\\
& & & & & \\
\hline
\end{tabular}
}
\end{center}
\centerline{\bf Table 2}

\end{document}